\def\astroph{1}                              % astro-ph or not
\ifnum\astroph=0                             % ApJL submission
\documentclass[12pt,preprint]{aastex}        % Single column, single spaced

\else                                        % ApJL emulation, for astro-ph
\documentclass{emulateapj}
\usepackage{apjfonts}
\fi

\usepackage{epsfig,graphicx,natbib}
\usepackage{subfigure}
\newcommand{\fig}[1]{Fig.\ \ref{#1}}
\newcommand{\Fig}[1]{Figure \ref{#1}}

\slugcomment{Accepted in \apj L}

\begin{document}
\author{Jacob Trier Frederiksen$^1$, Troels Haugb\o lle$^1$, Mikhail V. Medvedev$^{1,2,3}$, \AA ke Nordlund$^1$}
\affil{$^1$Niels Bohr International Academy, Blegdamsvej 17, DK-2199 K\o benhavn \O, Denmark}
\affil{$^2$Institute for Advanced Study, School of Natural Sciences, Princeton, NJ 08540}
\altaffiltext{3}{Also at: Department of Physics and Astronomy, University of Kansas, Lawrence, KS 66045 and Institute for Nuclear Fusion, RRC ``Kurchatov Institute'', Moscow 123182, Russia}

\title{Radiation Spectral Synthesis of Relativistic Filamentation}
\email{trier@nbi.dk}

\shorttitle{Radiation Spectral Synthesis}
\shortauthors{Frederiksen, Haugb\o lle, Medvedev, Nordlund}


%-----------------------------------------------------------------
\begin{abstract}
Radiation from many astrophysical sources, e.g.\ gamma-ray bursts and active galactic nuclei, is believed to arise from relativistically shocked collisionless plasmas. Such sources often exhibit highly transient spectra evolving rapidly, compared with source lifetimes. Radiation emitted from these sources is typically associated with non-linear plasma physics, complex field topologies and non-thermal particle distributions. In such circumstances a standard synchrotron paradigm may fail to produce accurate conclusions regarding the underlying physics. Simulating spectral emission and spectral evolution numerically in various relativistic shock scenarios is then the only viable method to determine the detailed physical origin of the emitted spectra. In this Letter we present synthetic radiation spectra representing the early stage development of the filamentation (streaming) instability of an initially unmagnetized plasma, which is relevant for both collisionless shock formation and reconnection dynamics in relativistic astrophysical outflows, as well as for laboratory astrophysics experiments. Results were obtained using a highly efficient \textit{in situ} diagnostics method, based on detailed particle-in-cell modeling of collisionless plasmas. The synthetic spectra obtained here are compared with those predicted by a semi-analytical model for jitter radiation from the filamentation instability, the latter including self-consistent generated field topologies and particle distributions obtained from the simulations reported upon here. Spectra exhibit dependence on the presence --- or absence --- of an inert plasma constituent, when comparing baryonic plasmas (i.e.\ containing protons) with pair plasmas. The results also illustrate that considerable care should be taken when using lower-dimensional models to obtain information about the astrophysical phenomena generating observed spectra.
\end{abstract}

%----------------------------------------------------------------------
\section{Introduction}
High-energy astrophysical emission associated with localized and transient sources - for example gamma-ray bursts (GRBs) --- is believed to arise from relativistically shocked plasmas. The transition from up-stream unshocked plasma to down-stream shocked plasma involves a rapid decrease in Lorentz factor over the shock ramp, with counter-streaming plasmas and associated streaming instabilities as a consequence -- see i.e.~\cite{bib:Meszaros2002}. Likewise, in outflows from relativistic reconnection events, unsteady short lived collisionless shocks are formed by relativistic streaming instabilities (i.e. ~\cite{bib:ZenitaniHesse2008}). Reconnection is often associated with flaring events, for example in solar flares and in transients of GRBs (flaring/sub-bursts).

Streaming instabilities, including the filamentation instability (FI), are effectively at work in the propagation of collisionless shocks in initially non-magnetized, or weakly magnetized, regions. The instabilities result in the generation of complex and time-dependent electro-magnetic fields, with structures covering a range of scales. Detailed phase space and field topology information is needed in order to extract precise diagnostics from such flows,
where the standard synchrotron paradigm assumptions are explicitly violated ~\citep{bib:MedvedevLoeb99,bib:SilvaEtAl2003,bib:FrederiksenEtAl2004,bib:Spitkovsky2008}.

Particle-in-cell (PIC) models are naturally suited for this task, since radiation spectra may be synthesized directly from such simulations by simply integrating the expression for the radiated power, derived from the Li\'{e}nard-Wiechert potentials for a large number of representative particles in the PIC representation of the plasma~\citep{bib:HededalThesis2005,bib:PhoucEtAl2005,bib:NishikawaEtAl2008,bib:SironiEtAl2009}.

Shock plasma conditions develop through several stages, from two-stream non-thermal to almost isotropically thermal. All radiative stages of the relevant instabilities should be probed when modeling optically thin emission spectra from highly transient phenomena. Here we exemplify one such transition, passing through several radiative stages of the FI, with a PIC simulations of a spatially periodic and initially unmagnetized two-stream system.

We emphasize the exploratory nature of this setup, which provides a well defined and reproducible model setup where such a transition occurs selfconsistently. We make no claim that there is a one-to-one mapping of spatial features and time scales to real phenomena.

Section 2 outlines our synthesis method and PIC simulations while Section 3 is dedicated to results. Discussions and conclusions are found in Section 4.

\begin{figure*}[!t]
 \epsfig{figure=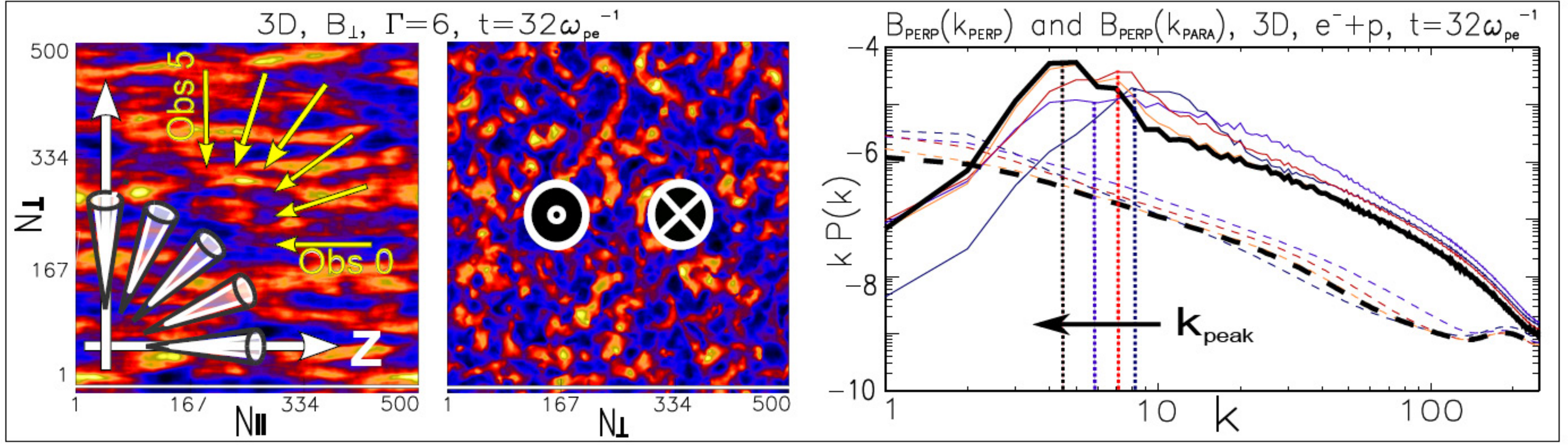, width=\textwidth}
 \caption{\textit{Left:} transverse magnetic field, $B_\perp(\vec{\perp},\vec{z})$ along the streaming direction. Yellow arrows show how multiple observers are distributed on $\angle[\vec{obs},\vec{z}]\in[0,..,\pi/2]$ w.r.t. the streaming direction, $\vec{z}$. Narrow white cones with opening angles $\delta\theta=1/\Gamma_{Bulk}$ indicate directions, $\theta_C\equiv\angle(\vec{z},\vec{cone})$, $\theta_C\in[0,...,\pi/2]$, of solid angle elements chosen in testing for anisotropy in momentum distribution and acceleration of electrons -- cf. section~\ref{sec:noniso_accel}. \textit{Middle:} transverse magnetic field, $B_\perp(\vec{x},\vec{y})$ across the streaming direction. Two identical plasmas (either $e^-+e^+$ or $e^-+p$) are counter-streaming, as indicated by the circular symbols. \textit{Right:} temporal development of magnetic field power spectra taken across ($\widetilde{B}_\perp(k_\perp)$ -- solid line) and along ($\widetilde{B}_\perp(k_\parallel)$ -- dashed line) the streaming direction $\vec{z}$. Black lines represents latest simulation snapshot, and colored lines indicate predecessive snapshots. Vertical dashed lines indicate $k_{peak}$ location.}
\label{fig:figure1final}
\end{figure*}

%============================================
\section{Simulations} \label{sec:simulations}
%-------------------------------------------------
\subsection{Filamentation Instability Simulations}
The relativistic filamentation instability is set up by preparing two equal density neutral plasma beams (\fig{fig:figure1final}), under fully periodic boundary conditions. Beams were chosen with bulk flows of $\Gamma_{Bulk}\in[2,4,6,10,15]$ ($\Gamma_{Bulk}\in[4,6,10]$) for 3 (2) spatial components and 3 (2) velocity components, or 3D3V (2D2V) runs. Simulation volumes were $500{\Delta}x\times500{\Delta}y\times500{\Delta}z=50^3\delta_e^3$ ($6000{\Delta}x\times6000{\Delta}z=300^2\delta_e^2$), respectively. Plasma densities were adjusted such that all simulations contained an equal number of relativistic electron skindepths, $\delta_e=\omega_{pe}^{-1}c$, independent of $\Gamma_{bulk}$, with $\omega_{pe}=[4{\pi}n_eq^2/m_e\Gamma_{bulk}]^{1/2}$ the relativistic plasma frequency. We simulated both baryonic ($e^-+p^+$) and pair ($e^-+e^+$) plasmas. A reduced mass ratio was used for the baryonic case; $m_i/m_e\equiv64$ suffices in separating ion and electron dynamic time scales.

In addition, a subset of the radiation spectra were collected in static (frozen) EM fields, starting from a well developed time in the simulation, rather than in \emph{in situ} full electromagnetic dynamics. These 'frozen' test runs are analyzed and discussed by~\cite{bib:MedvedevEtAl2010}. In summary they conclude that, unlike realistic synthetic spectra collected \emph{in situ}, spectra obtained in static (yet self-consistent) fields are well (but erroneously) interpreted by a synchrotron template. This, along with the obvious advantage of computational speed and ease was a main motivation for employing a completely self-consistent \emph{in situ} synthesis approach.

We chose the often employed PIC scaling of natural constants, setting $c{\equiv}m_e{\equiv}q_e{\equiv}1$. Varying $n_e$ (number or charge density) ensured an equal number of skindepths for all $\Gamma_{Bulk}$ for reasons of comparison. Time is henceforth normalized to $t\omega^{-1}_{pe}\equiv1$. All results were obtained using a 6$^{th}$ order field solver for Maxwell's equations. The field spectra were only affected qualitatively at high wavenumber (i.e. in the extreme dissipative range). Spectral synthesis was verified to exhibit negligible dependence of field solver order. We used cubic (quadratic) spline interpolation between particles and the field grid in 3D (2D).

%----------------------------------------
\subsection{Radiation Spectral Synthesis}\label{sec:radspecsynth}
We incorporate into the \textsc{PhotonPlasma} PIC code, developed in Copenhagen~\citep{bib:HaugboelleThesis2005,bib:HededalThesis2005}, a discretized version of the standard expression for radiated power from accelerated charges $P(\omega){\propto}d^2W/d{\Omega}d\omega$ (e.g.~\cite{bib:RybickiLightman}). Using the aforementioned PIC scaling, synthetic radiation frequencies are normalized to $\omega_{pe}$. Spectra are collected at runtime, and for simulations using $N_{tot}{\sim}10^{10}$ particles, we collect spectra from $N_{synth}{\sim}10^6$ particles without noticeable slow-down in simulation speed. We retain all phase and temporal information of the ensemble signal by integrating the signal as:
\[P_\omega\propto|\int^{t1}_{t0}\sum_j^{N_{synth}}E_{ret,j}e^{i\omega\phi'}dt'|^2,\] with $\phi'{\equiv}t'-\vec{n}\cdot\vec{r_0}(t')/c)$.  Spectra are recorded every small multiple of a simulation timestep. Thus, we may extract 'instantaneous' spectral states as $P_\omega(t_1,0)-P_\omega(t_0,0)$. We can specify multiple (far-field) observers at arbitrary orientation, and we can choose multiple regions for spectral collection in the simulation volume. In practice, we have a choice of selecting phase coherent or phase incoherent spectral synthesis. Binning over frequency spectral range was chosen to be logarithmic in $\Delta\omega$, but can in general be either logarithmic or linear in $\Delta\omega$; linear frequency binning is used primarily for test purposes. Particles selected for synthetic radiation contribution are subcycled, typically ${\Delta}t_{synth}\leq 10^{-1}{\Delta}t_{sim}$ to resolve high frequency radiation emitted by particles with $\beta_{em}\approx1$ and $\vec{n}\parallel\vec{\beta}$. Six observers (placed at 'infinity') in 3D (2D) are spread uniformly over angles, $\angle\{\vec{z},\vec{obs}\}\in[0,\pi/2]$ with respect to the streaming direction. Observer "0" designates head-on, and observer "5" is edge-on (3D case).

A number of plasma effects are neglected here. They shall be addressed in future work: plasma frequency cutoff, Razin effect and self-absorption processes. The spectral synthesis method has been sanity checked exhaustively~\citep{bib:HededalThesis2005,bib:HededalNordlund2005}.

\begin{figure*}[!ht]
  \epsfig{figure=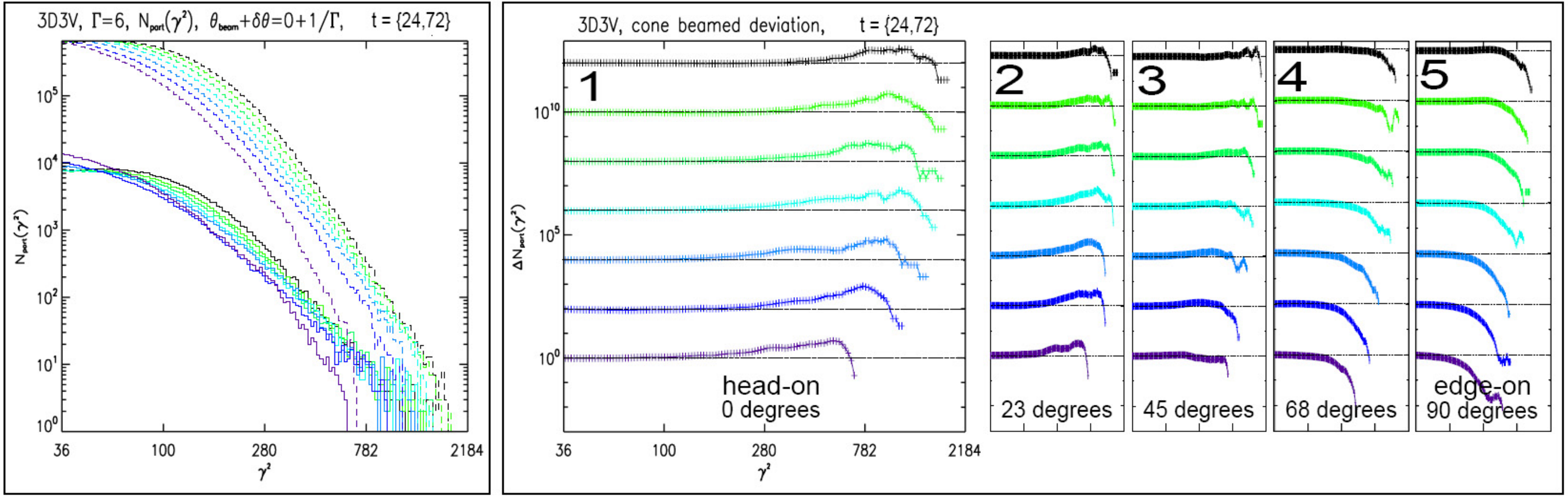, width=\textwidth}
  \caption{\textit{Left panel:} total particle distribution (dashed curves) in $\gamma^2$, above $\gamma^2=\Gamma^2_{Bulk}$. Particle distribution subset (solid curves) inside a narrow cone, $\angle\{\vec{p_z},\hat{z}\}\lesssim\pm\Gamma_B^{-1}$, pointing head-on the streaming direction, i.e. $\theta_C\equiv0$. \textit{Right panel:} inset "1" plots \textit{relative deviation from isotropy} of the sub-population-to-total population. Curves are offset by factors of $\times10$ for clarity. Color indicates time; purple (bottom) = earliest, black (top) = latest. Relative deviation is defined such that $\Delta N(\gamma^2)=4=3+1$ means three times more particles at given $\gamma^2$ than is present assuming an isotropic distribution. The 4 narrow panels (insets "2"-"5") view the same \textit{relative} deviation from isotropy now for cones, $-\Gamma_B^{-1}\leq\angle\{\vec{p_z},\hat{z}\}-(m\pi/8)\leq +\Gamma_B^{-1}$ for $m=\{1,\ldots,4\}$, oriented at successively inclined angles (indicated by white cones in \fig{fig:figure1final} leftmost panel) w.r.t. the streaming direction. Color coding designates simulation time, with $t\in[24, 32,\ldots,72]$ (bottom-to-top), respectively, in all panels. Plots are for the 3D baryonic ($e^-+p^+$) case. All plots are Log-Log. \textit{NB;} This analysis does not assume Maxwellians for the (total) isotropic ensemble. We can still have isotropy with non-Maxwellians -- which is seen to be the case; even the total ensemble is non-thermal.}
\label{fig:anisotrop_accel}
\end{figure*}
%===================================================
\section{Results} \label{sec:results}
%-------------------------------------------------------------------------------
\subsection{Evolution of the Filamentation Instability} \label{sec:instability_evolution}
Growth rates of both linear and non-linear FI depend heavily on the plasma constituents~\citep{bib:FrederiksenEtAl2004}; in $e^-+p$ plasmas, due to the presence of an inert species, growth rates are reduced~\citep{bib:TzoufrasEtAl2006}, compared with $e^-+e^+$ plasmas. Emerging electromagnetic field topologies and power spectra are likewise affected. For our setup two intervals of instability, and a brief turn-over, are identified for the relativistic FI: 1) linear, $t\in[0,12]$ ($t\in[0,14]$, for $e^-+e^+$ case), 2) saturation/transition, $t=[14,16]$ ($t\in[12,14]$), and 3) non-linear stage, $t\in[16,$"$\infty$"$]$ ($t\in[14,$"$\infty$"$]$). We shall use these intervals when comparing spectra from baryonic and pair plasmas. (electro-)Magnetic fields develop rapidly during the linear phase. Subsequent to saturation the field spectra converge to a quasi-stationary configuration. A representative case is plotted in the rightmost panel of~\fig{fig:figure1final} for late times of the evolution. The perpendicular Fourier spectra of the magnetic field, $\widetilde{B}_{k_\perp}\equiv\langle\widetilde{B}_{\perp}({k_\perp})\widetilde{B}^*_{\perp}({k_\perp})\rangle$, fit a Gaussian + a high frequency powerlaw segment on the upper dynamic range, with a dissipative cutoff for $k{\rightarrow}k_{Ny}$ (Nyquist scale). 'Parallel' Fourier spectra of the magnetic field, $\widetilde{B}_{k_\|}\equiv\langle \widetilde{B}_{\perp}({k_\|})\widetilde{B}^*_{\perp}({k_\|})\rangle$, are single powerlaws on $k_{\parallel,0}<k_\parallel<k_{\parallel,Ny}$. Except for the earliest linear FI, temporal evolution of $\widetilde{B}_{\perp}$ is modest -- see \fig{fig:figure1final} (right panel). Particle PDFs are well represented by two shifted Gaussians in the streaming direction and a single Gaussian perpendicular to the streaming direction, $f(p_z,p_\perp,\Gamma_B)\sim\left(exp(-(p_z-\Gamma_B)^2)+exp(-(p_z+\Gamma_B)^2)\right)exp(-p_\perp^2)$. The average Lorentz factor of the streaming particles decrease in the simulation (lab) frame during the FI. Temperatures increase with $T_\perp>T_\parallel$ for early times and $T_\perp{\approx}T_\parallel$ in the late phase. Once the FI has burned all its momentum anisotropy, the total electron ensemble has $\Gamma_{bulk}\approx1$ (no bulk flow).

%-----------------------------------------------------------------------
\subsection{Non-isotropic electron acceleration during FI} \label{sec:noniso_accel}
%We analyzed momentum space for possible deviations from isotropy.
Non-isotropic momentum distributions naturally result from non-isotropic particle acceleration; we expect this to be the case based on our devised toy model of "non-Fermi" acceleration, ~\citep{bib:HededalEtAl2004,bib:Medvedev2006.1}.

We therefore proceeded to compare subsets of the total particle ensemble, in momentum space, specified by the criteria: \[\theta_{C,m}-\delta\theta_C~\leq~\arctan\left({|\vec{p_{\perp}}|^{-1}}{\vec{p_\parallel}}\right)~\leq~\theta_{C,m}+\delta\theta_C,\] effectively selecting particles with momenta confined inside a narrow pitch cone of opening angle $\delta\theta_C\equiv 1/\Gamma_{Bulk}$, at varying inclination, $\theta_{C,m}{\equiv}m\pi/8$, $m\in\{0,1,\ldots,4\}$, w.r.t. the streaming direction. Thus, electrons are selected which are significantly beaming their radiative contribution in the direction of the cone; they radiate efficiently for certain observer orientations, $\theta_C$.

\Fig{fig:anisotrop_accel} (insets 1 \& 2) show how, early during the FI, electrons experience a \textit{non-isotropic acceleration}, with directional preference towards forward beaming of $1/\Gamma_{bulk}$. This 'beamed' acceleration continues to drive a small part of the total particle ensemble to high gamma factors ($\gamma_{e,max}\sim45$). Later, as $\Gamma_{bulk}$ decreases (bi-directionally), this anisotropy spreads to include also higher inclination angles (inset 3). This could be either due to early 'preference-accelerated' electrons diffusing out of their streaming directed pitch cone (inset 1$\rightarrow$inset 4), or due to current filaments going unstable and acquiring a cross-stream component (i.e. they start bending), or both.

%-------------------------------------------------------------------------------------------
\subsection{Synthetic radiation, spectral evolution} \label{sec:synthspectra}
\begin{figure}[!hb]
    \epsfig{figure=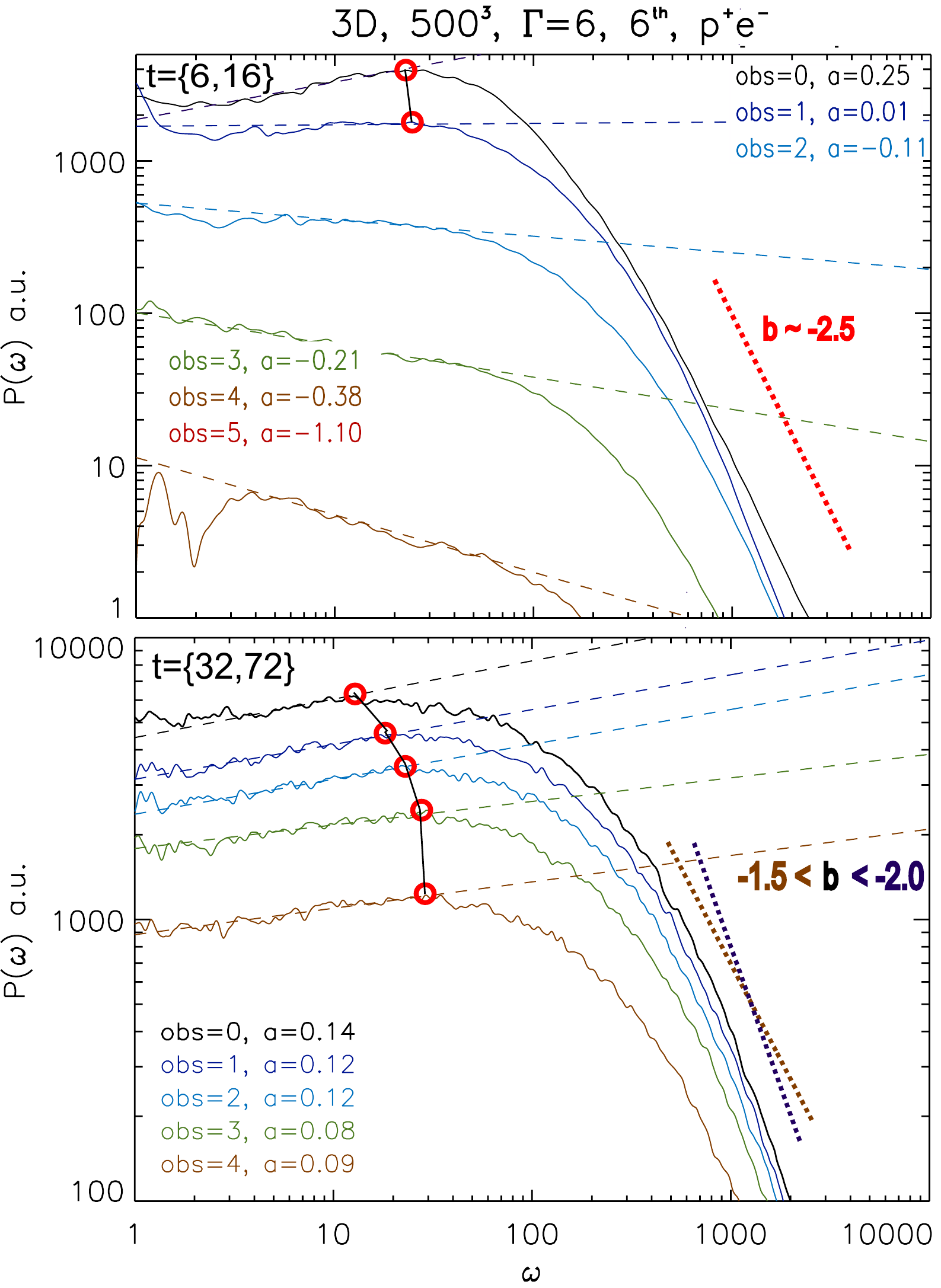, width=.48\textwidth}
    \caption{3D: Spectra for multiple observers. \textit{Upper panel:} Early stage FI. Spectra have barely had time to form. \textit{Lower panel:} Same but for late stage FI. Open circles (red) added, suggesting variation of peak frequency with observer angle. Dashed and dotted lines suggest low and high frequency powerlaw slope indices, $\alpha$ ("a") and $\beta$ ("b"), respectively. $\beta$ indicates range of possible slopes.}
\label{fig:3d6gFig3_ions_int2}
\end{figure}

Figure~\ref{fig:3d6gFig3_ions_int2} reveals consistently higher peak frequencies, $\omega_{peak}$, for increasing observer angles \textit{up to a maximum angle} $0<|\theta_{peak}|<\pi/2$. For higher angles $\omega_{peak}$ stays approximately constant up to $\theta\sim\pi/2$. Even at late times (\fig{fig:3d6gFig3_ions_int2}, lower panel) peak frequencies $\omega_{peak}$ are observer dependent, and apparently \emph{increase} with streaming axis inclination angle. Still, even though $\omega_{peak}(|\theta|>0)$ increases relative to the head-on ($\theta\equiv0$) observer at a given time, it never exceeds the initial head-on maximum peak frequency, $\omega_{peak}(\theta\equiv0)$, in 3D.

\begin{figure}[!ht]
    \epsfig{figure=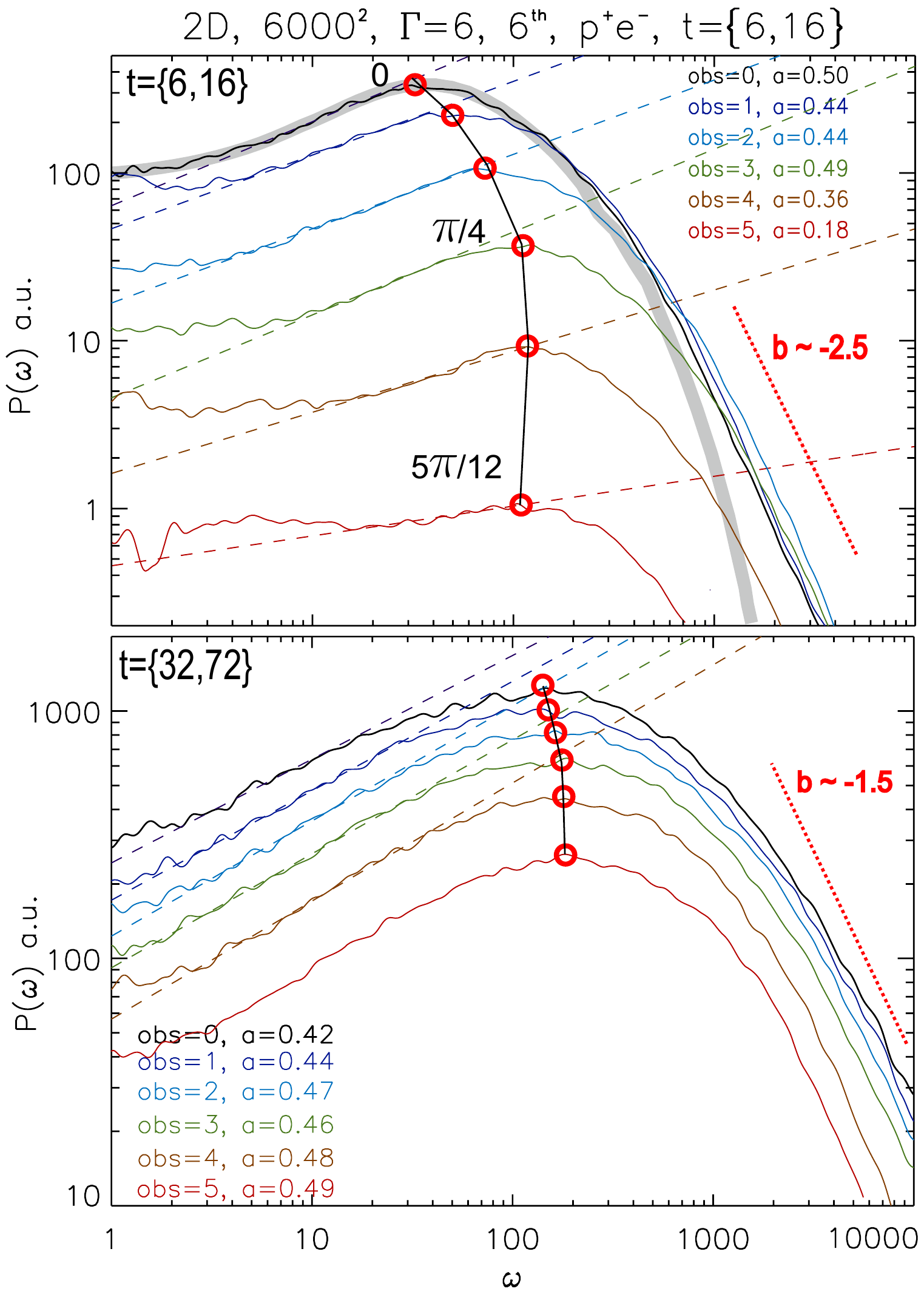, width=.48\textwidth}
    \caption{2D: Spectra for multiple observers. \textit{Upper panel:} early stage FI. Open circles (red) and thin broken line (black) added to suggest the variation of $\omega_{peak}$ with observer angle.
The off-axis variation of $\omega_{peak}$ is more pronounced in 2D; the probability that particles \emph{do not} pass through a filament is much smaller than for 3D. Additionally is given a "by-eye" fit (grey thick line), by realignment of the semi-analytical model (cf.\ the true 3D fit ~\fig{fig:3d6gFig1}). Dashed and dotted lines have been added to suggest low and high frequency powerlaw slope indices, $\alpha$ ("a") and $\beta$ ("b"), respectively. \textit{Lower panel:} Late stage FI. High frequency slopes are consistently $\beta\sim-1.5$ which is harder than 3D for all observers except head-on (cf.\ \fig{fig:3d6gFig3_ions_int2}). This is consistent with a 'filament hit probability' argument (cf. sect.~\ref{sec:synthspectra}).}
\label{fig:2d6gFig3_ions_early_alt1_noisecorr}
\end{figure}

In the 2D case viewed in \fig{fig:2d6gFig3_ions_early_alt1_noisecorr} the peak frequency increases for all observers during the FI, and $\omega_{peak, 2D}>\omega_{peak, 3D}$ for all cases (open circles of both \fig{fig:3d6gFig3_ions_int2} and \fig{fig:2d6gFig3_ions_early_alt1_noisecorr}). This effect is an artifact of the lower-dimensional 2D model; particles experience more frequent interactions with the ion filaments, since the "filament hit probability" is considerably smaller in 3D due to the reduction in filament filling factor (filaments are actually 'sheets' in 2D, whereas they are true line-like filaments in 3D). Radiation spectra are therefore formed more rapidly by the plasma field structures and with better statistics in 2D -- albeit artifactually so.

\begin{figure}
    \epsfig{figure=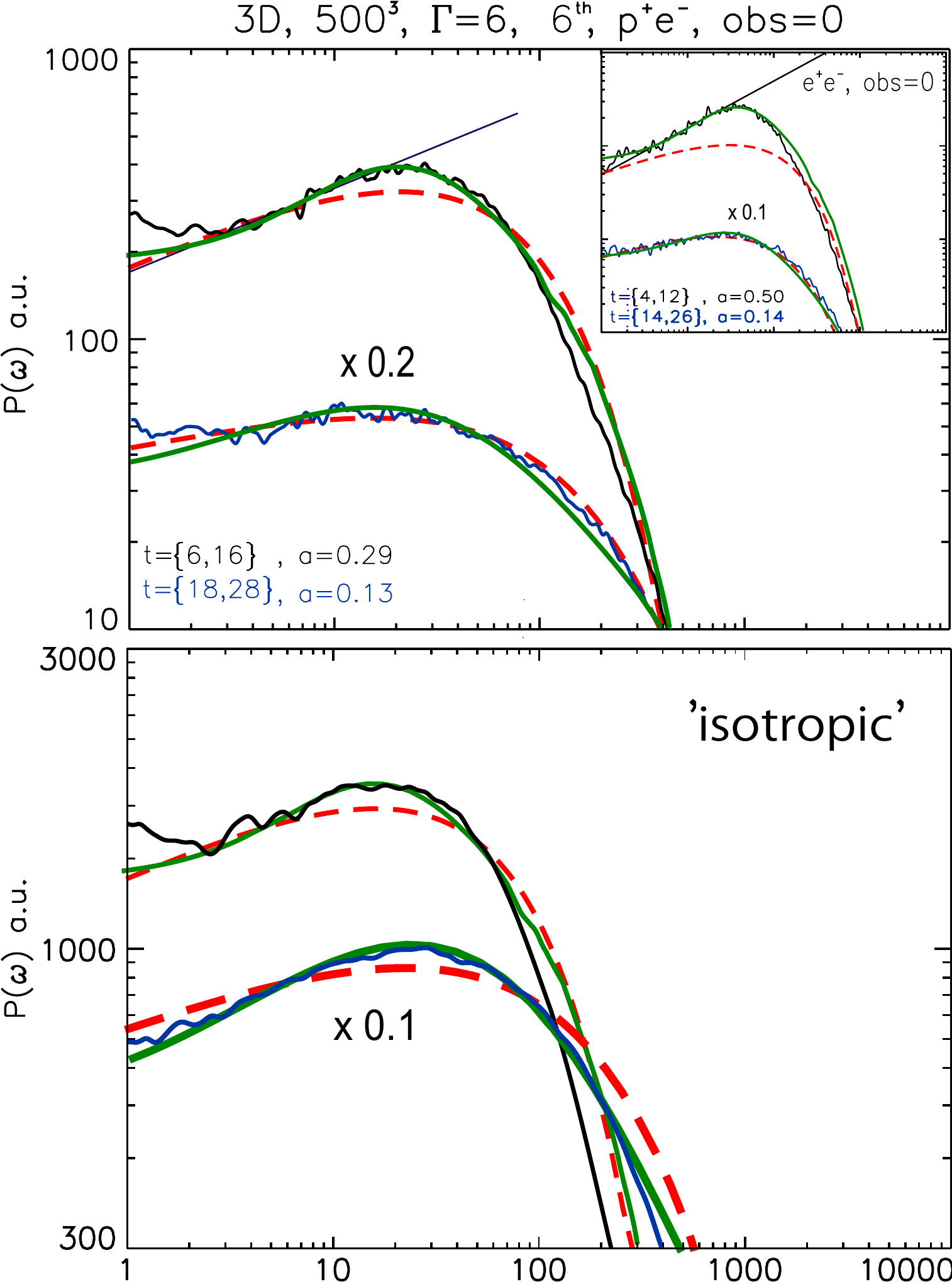, width=.48\textwidth}
    \caption{3D, baryonic plasma, head-on observer. \textit{Upper panel:} comparison of synthesized spectra with semi-analytical model fits to jitter (green, solid) and synchrotron models (red, dashed). Spectra for $t\in[6,16]$ and $t\in[18,28]$, for linear and late stages of the FI, respectively. Late stage is offset by factor of ${\times}0.2$ for clarity. NB; the semi-analytical fit in the baryonic case is "by-eye fit" of that for the pair plasma case, but realigned. The functional form still provides the better fit of the two models (jitter vs. synchrotron). Inset: same fits as in baryonic case, but for a pair plasma, with late stage spectrum offset by $\times 0.1$. This strict theoretical fit aligns remarkably well with the jitter case. \textit{Lower panel:} same as upper panel except for an isotropically averaged spectrum based on 6 observers at inclination angles w.r.t. the streaming direction as described in section~\ref{sec:radspecsynth}.}
\label{fig:3d6gFig1}
\end{figure}

Fitting the semi-analytical result for jitter radiation \citep{bib:MedvedevEtAl2010} to 3D spectra (\Fig{fig:3d6gFig1}, upper panel) shows markedly better fits than a synchrotron spectrum for the same parameters, during early stage FI, both for baryonic plasmas and pair plasmas. At later stages, also models of synchrotron fit the spectra well, but we emphasize previous findings, that the underlying plasma dynamics is not synchrotron in origin~\cite{bib:Medvedev2000, bib:Medvedev2006}. For a spectrum (and evolution) which has been isotropically averaged over all observer directions --- \Fig{fig:3d6gFig1}, lower panel --- making the same qualitative scaled fits of the semi-analytic model produces better agreement if a jitter radiation environment is assumed. Such an averaged spectrum can be seen as the immediate spectral state in the bulk of the plasma two-stream (or locally in a shock), when the simulation frame coincides with the co-moving frame of the plasma, as is the case in the FI setup presented here.

Our result further agrees quite precisely with the scaling relation of peak frequencies for a semi-analytical jitter radiation model, reported in ~\citep{bib:MedvedevEtAl2010}. At early times, we have measured $\omega_{peak,3D}\propto\Gamma^2_{bulk}$, while at late times $\omega_{peak, 3D}\propto\Gamma^3_{bulk}$ joint by a smooth transition.

A 'by-eye' re-alignment fit (thick grey line, \fig{fig:2d6gFig3_ions_early_alt1_noisecorr}) of the same semi-analytic result to the 2D baryonic case shows an under-estimate of high frequency radiated power. This undershoot artifact is not seen in 3D (upper green, solid, curve, \fig{fig:3d6gFig1}). It, too, is probably a consequence of the reduced dimensionality, again by the above argument concerning 2D result artifacts.

Moreover, in 2D, the maximum value corresponding to the abscissae of \fig{fig:anisotrop_accel},  $\gamma^2_{max, 3D}\approx2100$, is $\gamma^2_{max, 2D}\approx4200$ which is higher by a factor of $\gamma^2_{max, 2D}/\gamma^2_{max, 3D}\sim2$. This further influences late time spectra (lower panels of \fig{fig:3d6gFig3_ions_int2} and \fig{fig:3d6gFig1}). The 2D and 3D cases exhibit gross discrepancy in peak frequencies; they differ by an order of magnitude ($\omega_{peak,2D}/\omega_{peak,3D}\approx10$), due to a general upward drift of $\omega_{peak}$ in 2D. Again, 2D spectra are pushed to higher frequencies, likely due to an enhanced 'filament hit probability'. The 2D results for $e^++e^-$-plasmas exhibit analogous features.

%===========================================================
\section{Discussion and Conclusion} \label{sec:conclusions}
The context of the current investigation is to compare the jitter and synchrotron radiation approaches to radiation spectra modeling in a simple but well defined and reproducible setup.  The more general problem of synthesizing the radiation field from fully developed collisionless shocks, with a firm tuning of parameters, is the subject of future studies.

Here we have produced synthesized radiation spectra for multiple observer orientations from 2D and 3D particle-in-cell simulations of the relativistic filamentation instability in a periodic box, relevant to relativistic collisionless shocks such as believed at work in for example gamma-ray bursts, and in transient shocks at reconnection sites. We investigated radiation spectral evolution for both pair plasmas ($e^-+e^-$) and baryon loaded ($e^-+p$) plasmas, for a broad range of bulk flow speeds.

In summary the results are:
\begin{itemize}
    \item At early times, $\omega_{peak,3D}\propto\Gamma^2_{bulk}$, which fits a semi-analytical jitter radiation model~\citep{bib:MedvedevEtAl2010}. Later $\omega_{peak, 3D}\propto\Gamma^3_{bulk}$, with a smooth transition between the two regimes. We emphasize that this scaling is not synchrotron in origin.

    It may be difficult to extract such scaling-laws directly from existing observations. However, the result is still quite relevant: through a shock interface, the bulk Lorentz factor and spectral parameters vary drastically. In employing semi-analytical models based on synchrotron modeling, inference of a most likely physical scenario might fall short, yielding incorrect scalings of frequencies and flux.

    \item Non-isotropic acceleration occurs during both linear and early non-linear FI, and is present in both baryonic and pair plasmas. For pairs the process progresses much faster, and non-isotropic acceleration is negligible. The findings are commensurate with previous findings on direct 'non-Fermi' particle acceleration~\citep{bib:HededalEtAl2004} in non-linearly developed $e^-+p$ (i.e. inert) filaments~\citep{bib:FrederiksenEtAl2004}.
   \item Synthetic spectra generated from 2D PIC simulations exhibit an upward drift of $\omega_{peak,2D}$ which is not seen in 3D -- cf.\ \fig{fig:3d6gFig3_ions_int2} and \fig{fig:2d6gFig3_ions_early_alt1_noisecorr}.
    \item Further, generally, 2D spectra differ qualitatively and quantitatively from 3D spectra in both $\omega_{peak}$, $P_{peak} \equiv P(\omega_{peak})$ and in spectral slopes, $\alpha$ ($\beta$) at low (high) frequency -- see \fig{fig:3d6gFig3_ions_int2} and \fig{fig:3d6gFig1}.
\end{itemize}

We find that lower-dimensional models (such as for example 2D2V or 2D3V) provide distorted measurements of parameters central to radiation modeling. Peak frequencies are generally too high, and low- (high-) frequency spectral slopes, $\alpha$ ($\beta$), are too steep (flat) compared with spectra obtained from 3D modeling. We expect that 2D3V (2D spatial dimensions, 3D particle velocity components) simulations will display properties intermediate of 2D2V and 3D3V.

While obtaining synthetic spectra from both pair plasmas ($e^++e^-$) and baryonic plasmas ($e^-+p$), we establish that spectra are qualitatively dependent on the plasma constituent species. This difference must be addressed when interpreting astrophysical observed spectra. We interpret the radiation spectra differences between baryon and pair plasma cases as a direct signature of screened current channel formation in the non-linear stage of the FI.

The presence of non-isotropically distributed electrons, accelerated preferentially in the streaming direction, could lead to spectra that peak at higher frequency than would be estimated from the \textit{isotropic} (thermal + powerlaw) particle ensemble. Since they are powerlaw distributed (left panel, \fig{fig:anisotrop_accel}) their contribution should also be powerlaw by origin. Although non-isotropic acceleration is also present in $e^+e^-$-plasmas it is negligible on long timescales, due to the lack of an inert plasma species. We conjecture that baryonic plasmas could therefore yield radiation signatures with harder powerlaw behavior at high-frequency ($\beta_{baryonic}>\beta_{pair}$), but softer spectral slope at low frequency ($\alpha_{baryonic}<\alpha_{pair}$). The electrostatic field, although screened, also plays a significant role for spectrum of emitted radiation. In the up-stream region ahead of the shock, where streaming instabilities are of early stage type, screened electrostatic effects are likely to produce heavily observer dependent effects. For global shock simulations with non-periodic boundary conditions such a non-isotropic acceleration mechanism will be sustained.

Our results establish tractability of an alternative approach to radiation modeling -- the jitter radiation approach. It contrasts the standard synchrotron approach; assuming a pure synchrotron mechanism in observed relativistic streaming phenomena, we would infer --- incorrectly --- that large scale homogenous magnetic fields were present, when rather it might be small scale fields, in (a priori) non-magnetized relativistic shocked outflows.

A tight correspondence exists between  on one hand the spectral evolution of radiation emitted from electrons in  optically thin collisionless plasmas and, on the other, the field spectral evolution in the plasma due to the instabilities  that create and mediate such radiation.

Care should be exhibited with reduced dimensionality modeling; as 3D models become increasingly affordable they are to be preferred over 2D models, as spectral diagnostics can be distorted as a consequence of reduced dimensionality---filament entanglement is for example only possible in 3D. Attention should also be given to choices of plasma constituents (electrons, protons, ions, etc.).

To conclude, the radiation spectral modeling results presented here reveal limitations that need to be considered while reverse engineering spectra observed from, for example, relativistic astrophysical outflows, GRBs, relativistic reconnection sites, and from laser-plasma interaction in laboratory astrophysics, when examining likely physical scenarios for their match to observed spectra.

%---------------------------------------------------------
\acknowledgements {\AA}N, JTF, and TH acknowledge support from the Danish Natural Science Research Council. MVMs work has been supported by NSF grant AST-0708213, NASA ATFP grant NNX-08AL39G and DOE grant  DE-FG02-07ER54940. MVM also acknowledges support from The Ambrose Monell Foundation (IAS) and The Ib Henriksen Foundation (NBIA). Computer time was provided by the Danish Center for Scientific Computing (DCSC). This work was supported by the European Commission through the SOLAIRE Network ({\AA}N).

%---------------------------------------------------------
\bibstyle{apj}


\begin{thebibliography}{99}

\bibitem[Frederiksen et al.(2004)]{bib:FrederiksenEtAl2004} Frederiksen, J.~T.,
Hededal, C.~B., Haugb{\o}lle, T.,
\& Nordlund, {\AA}.\ 2004, \apjl, 608, L13

\bibitem[Haugb\o lle (2005)]{bib:HaugboelleThesis2005}
{Haugb\o lle}, T. 2005, PhD thesis, Niels Bohr Institute [astro-ph/0510292]

\bibitem[Hededal et al.(2004)]{bib:HededalEtAl2004} Hededal, C.~B.,
Haugb{\o}lle, T., Frederiksen, J.~T.,
\& Nordlund, {\AA}.\ 2004, \apjl, 617, L107

\bibitem[{{Hededal}(2005)}]{bib:HededalThesis2005}
{Hededal}, C. 2005, PhD thesis, Niels Bohr Institute [astro-ph/0506559]

\bibitem[Hededal \& Nordlund (2005)]{bib:HededalNordlund2005}
Busk Hededal, C., \& Nordlund, {\AA}.\ 2005, arXiv:astro-ph/0511662

\bibitem[Medvedev \& Loeb (1999)]{bib:MedvedevLoeb99}
Medvedev, M. V.; Loeb, A., \apj, vol.526, 697-706 (1999)

\bibitem[Medvedev(2000)]{bib:Medvedev2000} Medvedev, M. V.\ 2000, \apj, 540, 704

\bibitem[Medvedev(2006a)]{bib:Medvedev2006} Medvedev, M. V.\ 2006, \apj, 637, 869

\bibitem[Medvedev(2006b)]{bib:Medvedev2006.1} Medvedev, M. V.\ 2006, \apjl, 651, L9

\bibitem[Medvedev et al.(2010)]{bib:MedvedevEtAl2010} Medvedev, M. V., Frederiksen, J. T., Haugb\o lle, T. \& Nordlund \AA, 2010, submitted to \apj, preprint at arXiv:1003.0063v2  [astro-ph.HE]

\bibitem[M{\'e}sz{\'a}ros(2002)]{bib:Meszaros2002} M{\'e}sz{\'a}ros, P.\ 2002, \araa, 40, 137

\bibitem[Nishikawa et al. (2008)]{bib:NishikawaEtAl2008}
Nishikawa, K.~I., et al.\ 2008, conf. proc.,"Blazar Variability across the Electromagnetic Spectrum", Palaiseau, France (2008)

\bibitem[Phouc et al. (2005)]{bib:PhoucEtAl2005} Ta Phuoc, K., Burgy, F., Rousseau, J.-P. , Malka, V., Rousse, A., Shah, R.,
Umstadter, D., Pukhov, A., \& Kiselev, S.\ 2005, Phys. of Plasmas 12, 023101

\bibitem[Rybicki \& Lightman (1979)]{bib:RybickiLightman}
Rybicki, G. B. \& Lightman, A. P., "Radiative processes in astrophysics", Wiley \& Sons, NY, 1979

\bibitem[Silva et al.(2003)]{bib:SilvaEtAl2003} Silva, L.~O., Fonseca, R.~A., Tonge, J.~W., Dawson, J.~M., Mori, W.~B., \& Medvedev, M.~V.\ 2003, \apjl, 596, L121

\bibitem[Sironi \& Spitkovsky(2009)]{bib:SironiEtAl2009} Sironi, L.,
\& Spitkovsky, A.\ 2009, \apjl, 707, L92

\bibitem[Spitkovsky(2008)]{bib:Spitkovsky2008} Spitkovsky, A.\ 2008,
\apjl, 673, L39

\bibitem[Tzoufras et al.(2006)]{bib:TzoufrasEtAl2006} M. Tzoufras, C. Ren, F. S. Tsung, J.W. Tonge, W. B. Mori, M. Fiore, R. A. Fonseca, L. O. Silva\ 2006, \prl, 96, 105002

\bibitem[Zenitani \& Hesse(2008)]{bib:ZenitaniHesse2008} Zenitani, S., Hesse, M.\ 2008, Phys. of Plasmas 15, 022101

%--------------------
\end{thebibliography}
\end{document}